\documentclass{emulateapj}
\usepackage{graphics}
\usepackage{amssymb}
\usepackage{amsmath}
\usepackage[dvips]{color}
\newcommand{\kms}{{km~s$^{-1}$}}
\def\2da{\mbox{2175  \AA~absorbers}}
\newcommand{\rev}[1]{{\color{black} #1}}

\begin{document}

\title{Seven broad absorption line quasars with excess broad band absorption near 2250 \AA}
\author{ Shaohua Zhang\altaffilmark{1,2}, Jian Ge\altaffilmark{2}, 
Peng Jiang\altaffilmark{3}, Hongyan Zhou\altaffilmark{1,3}, 
Jingzhe Ma\altaffilmark{2}, W. N. Brandt\altaffilmark{4}, Donald G. York\altaffilmark{5}, 
Pasquier Noterdaeme\altaffilmark{6}, Donald P. Schneider\altaffilmark{7,8}}
\altaffiltext{1}{Polar Research Institute of China, 451 Jinqiao Road, Shanghai, 200136, China; zhangshaohua@pric.org.cn}
\altaffiltext{2}{Department of Astronomy, University of Florida, Gainesville, FL, 32611, USA; jge@astro.ufl.edu}
\altaffiltext{3}{Key Laboratory for Research in Galaxies and Cosmology, University of Science and Technology of China, Chinese Academy of Sciences, Hefei, Anhui, 230026, China}
\altaffiltext{4}{Department of Astronomy and Astrophysics, 525 Davey Laboratory, Pennsylvania State University, University Park, PA, 16802, USA}
\altaffiltext{5}{Department of Astronomy, University of Chicago, 5640 South Ellis Ave, Chicago, IL 60637, USA}
\altaffiltext{6}{Institut d'Astrophysique de Paris, CNRS-UPMC, UMR7095, 98bis bd Arago, 75014 Paris, France}
\altaffiltext{7}{Department of Astronomy and Astrophysics, The Pennsylvania State University,University Park, PA 16802} 
\altaffiltext{8}{Institute for Gravitation and the Cosmos, The Pennsylvania State University,University Park, PA 16802}

\begin{abstract}
We report the discovery of excess broad band absorption near 2250 \AA~(EBBA)
in the spectra of seven broad absorption line (BAL) quasars.
By comparing with the statistical results from the control quasar sample, 
the significance for the detections are all above the $\gtrsim 4\sigma$ level, with five above  $>5\sigma$. 
The detections have also been verified by several other independent methods. 
The EBBAs present broader and weaker bumps at smaller wavenumbers than the Milky Way, and similar to the Large Magellanic Cloud.
The EBBA bump may be related to the 2175  \AA~bump seen in the Local Group 
and may be a counterpart of the 2175  \AA~bump under different conditions in the early Universe.
Furthermore, five objects in this sample show low-ionization broad absorption lines (LoBALs), such as \ion{Mg}{2} and \ion{Al}{3}, 
in addition to the high-ionization broad absorption lines (HiBALs) of \ion{C}{4} and \ion{Si}{4}.
The fraction of LoBALs in our sample, $\sim$70\%, is surprisingly high compared to that of general BAL quasars, $\sim$10\%. 
Although the origin of the bump is still not clear, the coexistence of both BALs and bumps and
the significantly high fraction of LoBALs may indicate 
the bump carriers is closely related to the early evolution phase of quasars.
\end{abstract}


\section{Introduction}
Absorption systems  identified in quasar spectra are classified  as ``quasar-associated'' or
``intervening'' absorbers based on the nature of absorbing materials. Compared to the origin of 
 intervening absorbers which are produced by  absorbing materials, such as intergalactic clouds or galaxies, 
located far from the background quasars, the origin of quasar-associated absorbers is much more 
complicated. While most quasar-associated absorbers are produced by materials associated with the AGN environments,
a significant fraction may also be associated with AGN outflows. 
Therefore, quasar-associated absorbers uniquely trace a variety of regions in AGNs and their host galaxies,
and their processes are closely connected to the basic physics of the central Super-Massive Black Hole (SMBH) growth
and AGN evolution (Hamann \& Sabra 2004 and references therein).
Broad absorption lines (BALs) are the most {\rev prominent} quasar-associated absorbers because of their high velocities
(up to a velocity of $v\sim0.2c$) and broad widths (at least 2000 \kms~wide; see Weymann et al. (1991)).

BALs appear in the spectra of $\sim10-15\%$ of optically selected quasars, and often show absorption from
a wide range of species, such as \ion{Mg}{2}, \ion{Al}{3} and \ion{Fe}{2}, to \ion{N}{5}, \ion{C}{4}, \ion{Si}{4} and
\ion{O}{6} (Hall et al. 2002; Tolea et al. 2002; Hewett \& Foltz 2003; Reichard et al. 2003; Trump et al. 2006; Gibson et al. 2009;
Zhang et al. 2010, 2014). BALs are  clearly formed by gas outflows from the AGN central engine.
The accretion of gaseous material onto the central SMBHs releases a large amount of
 radiative and kinetic energy to heat and expel the surrounding gas and  
accelerates the outflowing gas  to very high velocities, producing
very broad velocity profiles in the absorption lines (see Antonuccio-Delogu \& Silk 2010 for a recent review).
Besides the high velocities and broad line profiles,  BAL quasars 
show remarkably weak soft X-ray emission (e.g., Green \& Mathur 1996),
with $\sim10-30$ times less soft X-ray flux than that for unabsorbed quasars
(e.g., Brandt, Laor, \& Wills 2000). The apparent X-ray absorption probably implies intrinsic extinction in BALs.
Previous statistical studies of optically selected BAL quasars indicate that
High-ionization BAL (HiBAL) quasars as a population are not 
heavily reddened ($E(B-V)\sim0.03$) while Low-ionization BAL (LoBAL) quasars
show a moderate level of reddening with the average color excess of $E(B-V)\sim0.1$ (Weymann et al. 1991;
Sprayberry \& Foltz 1992; Reichard et al. 2003; Gibson et al. 2009; Zhang et al. 2010). 
{\rev Both HiBAL and LoBAL quasars often show intrinsic dust reddening following the Small Magellanic Cloud (SMC) type
as do intervening quasar absorption line systems (Reichard et al. 2003; York et al. 2006),} 
unlike the Milky Way (MW) type dust reddening which
possesses the remarkable and intriguing broad 2175 \AA~bump in the UV extinction curve (Stecher 1965). 
The bump strength increases gradually from the SMC type curve to the Large Magellanic Cloud (LMC) like curve to the MW type curve
(Savage \& Mathis 1979; Fitzpatrick 1989).

Detections of 2175 \AA~features in high redshift ($z>1$) quasar absorption line systems were first reported by Wang et al. (2004).
Since then, over 40 2175 \AA~bumps have been found in high redshift quasar metal absorption systems,
such as \ion{Mg}{2} and \ion{Ca}{2} (e.g., Wild et al. 2006;
Srianand et al. 2008; Noterdaeme et al. 2009, 2014; Zhou et al. 2010; Jiang et al. 2010a,b; 2011; Wang et al. 2012).
{\rev Despite the existence of some known 2175 \AA~ bumps in quasar absorption line systems, 
they are extremely rare (York et al. 2006).  They may be expected to be even more rare in regions near quasars 
because the carrier of the bump may not survive in the harsh UV environment close to the quasars.
Indeed, only 6 quasar-associated 2175 \AA~ absorbers are found. }
The measurements of metal absorption lines associated with \2da~suggest that the dust depletion
in these absorbers is similar to that measured in cold interstellar clouds of the MW
(e.g. Jenkins 2009 and references therein), which are characterized by cold, dense and primarily molecular gas
(Jiang et al. 2010a,b; Zhou et al. 2010; Wang et al. 2012). This result could indicate that the carriers in these 
high redshift \2da~are likely the same as those responsible for the UV bumps in the MW and LMC. 

Although the origin of the 2175 \AA~bump remains unclear 50 years after its discovery (Stecher 1965),
carbonaceous materials have long been proposed as a candidate; 
a probable carrier is polycyclic aromatic hydrocarbon (PAH) molecules,
the suspected source of spectral features of which are abundant in the MW (Draine 2003; Wang et al. 2005).
If this is the case, then the 2175 \AA~bump can be detected in the quasar environment. 
For instance, the BAL winds can efficiently absorb the high-energy photons from the central engine of quasars
to allow survival of the PAH molecules or other carriers. Extreme ultraviolet and X-ray photons can easily destroy
PAH molecules (Voit 1992;  Murray et al. 1995). Detections of 
the unidentified emission signatures in the mid-infrared spectra of six Iron Low-Ionization BAL (FeLoBAL) quasars 
imply that carbonaceous molecules can survive in the environments of BAL quasars (Farrah et al. 2010).
However, there is no previous direct evidence of the coexistence of BALs and \2da~in quasar spectra.

In this paper, we report the discovery of seven possible \2da~in the optical spectra of BAL quasars
detected in the Sloan Digital Sky Survey III (Eisenstein et al. 2009).
The spectral analysis results are described in \S2, discussions about the false detections are presented in \S3 and 4
and conclusions and discussions are summarized in \S5.

\section{Spectra and Analysis}
The spectra of these seven objects were observed in the Baryon Oscillation Spectroscopic Survey (BOSS; Dawson et al. 2013)
of the third stage of the Sloan Digital Sky Survey (SDSS-III; Eisenstein et al. 2011) on a dedicated 2.5 m telescope (Gunn et al. 2006)
at Apache Point Observatory (APO) in the Sacramento Mountains in Southern New Mexico.
To constrain baryon acoustic oscillations (BAO), BOSS is designed to complete a spectroscopic survey of
 a larger volume than all of the 
previous spectroscopic surveys, including 1.5 million luminous galaxies as faint as $i=19.9$
with redshifts $z<0.7$ and more than 150,000 quasar spectra ($g<22$) over the redshift range $2.15<z<3.5$.
The improved BOSS spectrographs (Smee et al. 2013) can observe spectra of targets  two magnitudes fainter than the
 original SDSS spectrographs.
The calibrated digital spectra cover the wavelength region of 3,600-10,500 \AA~at a spectral resolution
in the range of $1,300<R<2,500$. 

\subsection{Searching procedures }
{\rev The quasar-associated \2da~are rarely detected, }
and the generally steep extinction curve (Fynbo et al. 2013; Jiang et al. 2013) makes some objects fainter.
Thus the BAL quasars with quasar-associated \2da~are rarer due to the 
additional dust reddening in the BAL clouds and their small percentage in the quasar population. 
Nevertheless, the improved survey sensitivity of BOSS allows detection of 
fainter objects with high-redshift in a larger volume than the original SDSS, 
offering a much larger sample of high redshift quasars to search for rare \2da. 
Our initial exploration of \2da, among \ion{Mg}{2} absorption line systems on quasar spectra in the SDSS DR10 (Ahn et al. 2014),
led to identification of excess broad band absorption near 2250 \AA~ in seven BAL quasars.

{\rev Searching procedure of the bump exploration has three steps, and we briefly outline it here.
Firstly, we construct a sample of \ion{Mg}{2} absorbers from the SDSS DR10 quasar catalog (P{\^a}ris et al. 2013).
\ion{Mg}{2} absorbers are identified by a semi-automatic algorithm  following the method of Lawther et al. (2012).
\ion{Mg}{2} absorbers have been visually inspected and some unusual absorption lines like those in Hall et al. (2002) are removed.
Based on the wavelength coverage of the BOSS spectrographs and the color distribution of \2da\ found by Jiang et al. (2011),  
7950 strong \ion{Mg}{2} absorbers with absorption redshifts of $0.7<z_{\rm abs}<3.0$ and relative colors of $\Delta (g-i) > 0$,
are collected as the parent sample of the exploration of \2da. 
Secondly, we use a quasar composite spectrum reddened by a parameterized extinction curve (introduced in \S 2.2)
to fit the observed spectra in the rest frame of absorption systems. 
This method is similar to the pair-method commonly used in studies of dust extinction of the MW diffuse clouds.
Thirdly, we perform an independent verification approach to gauge the significance of the extinction bumps 
using a simulation technique developed by Jiang et al. (2010a,b; also see the last paragraph of \S 2.4). 
The bumps with a significance level of $< 3\sigma$ are considered as false candidates to be rejecting.
After performing a visual examination and rejecting the false signals mimicked by greatly trimmed in the SDSS spectral data, 
we finally obtain 225 2175 \AA\ absorber candidates. Among them, 25 cases are quasar-associated 2175 \AA\ absorbers, 
and we find the absorption troughs of \ion{C}{4}BALs in the spectra of 7 objects.
Figure 1 shows the spectra of these seven BAL quasars, and the fitting results of bumps in BAL quasars are presented in \S 2.3.}

\subsection{Parameterization of extinction curves }
In the MW ISM studies, dust extinction curves of diffuse clouds are extracted 
by comparing a pair of stellar spectra of  the same spectral type, one of which is reddened and
the other unreddened (Fitzpatrick \& Massa 2007 and references therein). 
Since our technique to extract the extinction curves from quasar spectra is similar
to the pair-method, we denoted our technique the ``quasar spectrum pair method''.
In this method, the spectrum of the target is fit using the SDSS DR7 quasar composite spectrum
(Jiang et al. 2011) reddened by a parameterized extinction curve (Fitzpatrick \& Massa 1990)
at the optical/UV wavebands in the rest frame of the absorber of interest. 
The extinction curve $A(\lambda)$ is constituted by a linear background and a Drude component,
representing the underlying extinction and the potential 2175 \AA~bump, respectively.
\begin{align}
A(\lambda) 	&= c_1+c_2x+c_3 \dfrac{x^2} {(x^2-x_0^2)^2+x^2\gamma^2} \nonumber \\
			&= c_1+c_2x+c_3 D(x,x_0,\gamma),
\label{extinction}
\end{align}
where $x=\lambda^{-1}$ is in units of inverse microns ($\mu m^{-1}$). 
Two parameters ($c_1$ and $c_2$) define the linear background.
The Drude profile is defined by three parameters, (1) the position (in $\mu m^{-1}$) of the bump peak, $x_0$;
(2) the full width at half maximum (in $\mu m^{-1}$) of the bump, $\gamma$;
and (3) the Drude profile scale factor, $c_3$. 
The strength of the bump can be described by the area under the Drude profile, 
$A_{\rm bump} = \int_{0}^{\infty} c_3 D(x,x_0,\gamma)\, dx = \pi c_3/(2\gamma) 
A_{\rm bump} = E(B-V)\times A^{*}_{\rm bump}$, where $A^{*}_{\rm bump}$ is the bump strength 
defined in Fitzpatrick \& Massa (2007) and Gordon et al. (2003). Unlike the MW ISM studies, 
we do not know the intrinsic spectral slope and luminosity of the unreddened spectrum.
We cannot measure the conventional extinction parameters ($A_V$, $R_V$ and $E(B-V)$) 
from the extinction curve, derived by the comparison of the quasar spectra with the composite spectrum.
Thus, $A(\lambda)$ is a relative extinction curve without normalization.
Furthermore, the linear background also accounts for the variation of the intrinsic quasar spectral slopes and fluxes,
and $c_2$ could be negative if a quasar spectrum is bluer than the composite spectrum.
However, the aim of this work is to unveil the extinction bump in the quasar spectra,
{\rev and the features of the extinction bump will remain the same regardless of intrinsic quasar variability.}
More details can be found in Jiang et al. (2011).  

\subsection{Fitting results of bumps in BAL quasars}
In the fitting process of these seven BAL quasars, only the continuum of a quasar spectrum was fitted
while the regions with strong emission lines, such as Ly$\alpha$, \ion{C}{4}, \ion{Al}{3}, \ion{C}{3}
and \ion{Mg}{2}, and known strong BALs (gray curves in Figure \ref{f1}) are masked without fitting.
The potential excess broad band absorption is set to have the same redshift as the quasar's emission redshift to
detect the quasar-associated absorbers. We perform the least-squares minimization using 
the Interactive Data Language (IDL) procedure MPFIT developed by C. Markwardt.\footnote{The Markwardt IDL Library 
is available at http://cow.physics.wisc.edu/$\sim$craigm/idl/idl.html}
Figure \ref{f1} displays the spectra of the seven candidates with excess broad band absorption.
The best fit models (green solid lines) are presented with the SDSS-III/BOSS spectrum (black curves) in panel (a).
To emphasize the requirement of a bump on the extinction curve,
we also include the reddened composite quasar spectra with only the linear component of
the best fitting model (green dashed lines). The derived extinction curves are plotted in panel (a).
The best fit parameters of the extinction curves of the seven candidates are listed in Table \ref{tab1}
and shown in  panel (b) of Figure \ref{f1}. 
Figure \ref{f2} illustrates the distribution of the absorption feature parameters
and the relative extinction curves of the seven candidates.  
For comparison, Figure \ref{f2} also includes the bumps and the extinction curves
observed in the MW (Fitzpatrick \& Massa 2007; green circles and curves) and LMC (Gordon et al. 2003; red circles and curves). 
The peak position of the bump in the MW is nearly constant, 
and the bump width varies in a wide range of $\gamma=1.0\pm0.25~\mu m^{-1}$.
However, {\rev 7 cases in this work} present broader absorption features at smaller wavenumbers 
($\gamma=1.3\pm0.39~\rm \mu m^{-1}$, $x_0=4.45\pm0.05~\rm \mu m^{-1}$ or $2250\pm26$ \AA), 
and their bump strengths are much weaker than those of the MW bumps on average, but similar to those of the LMC bumps. 
The excess broad band absorbers are possibly related to the 2175 \AA~bump detected in our Milky Way and other galaxies.
The different distributions are possibly attributed to variations in chemical composition of the grains (Draine 2003).
For convenience, hereafter, we use the term excess broad band absorbers (EBBA) to refer to this new class of absorbers near 2250 \AA.

\figurenum{1}
\begin{figure*}[tbp]
\epsscale{1.0} \plotone{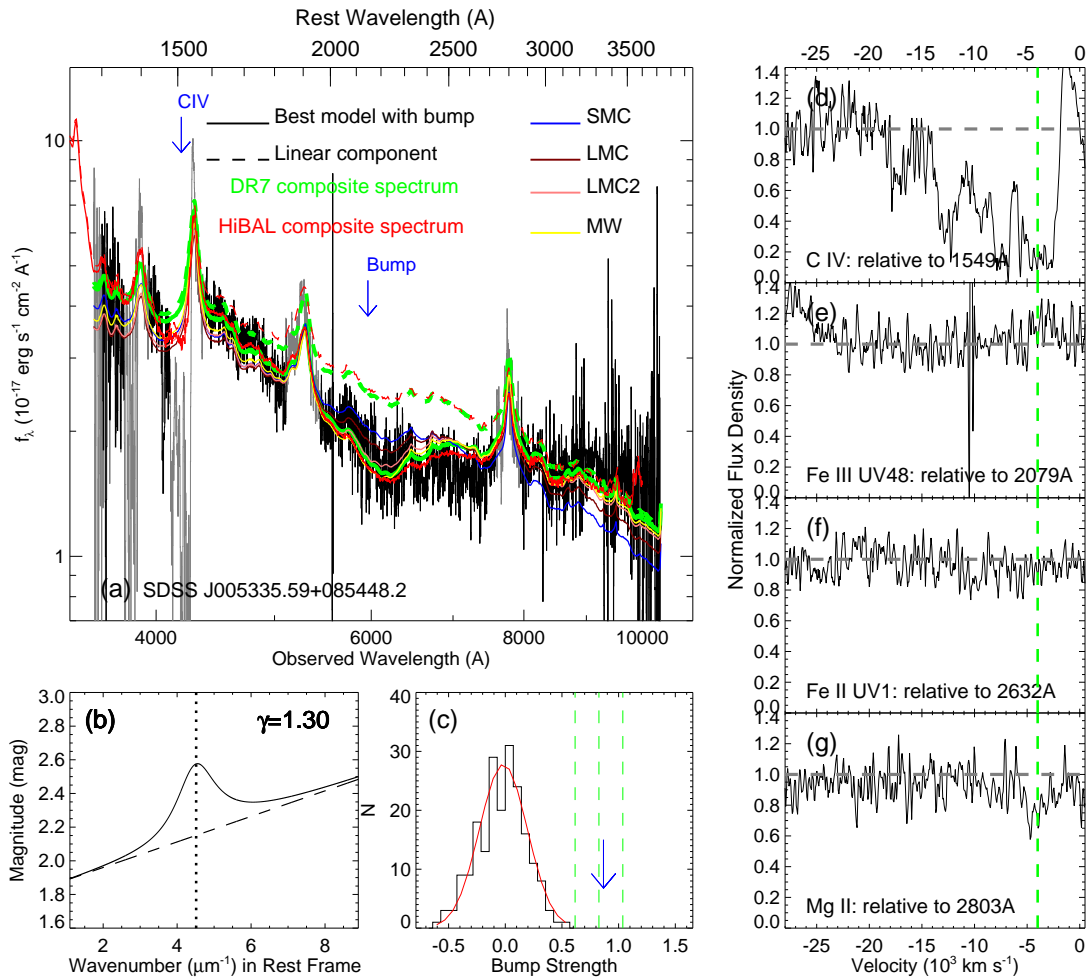}
\caption{ The best fit extinction model and absorption lines for J005335.59+085448.2.
(a). Green solid and dashed curves are the SDSS DR7 quasar composite spectrum reddened by the best parameterized model
with a linear component and a Drude component and with a linear component only, respectively. 
We also show the best fit results using the HiBAL quasar composite spectrum in Reichard et al. (2003) 
instead of the DR7 quasar composite spectrum in red.
By comparison, continuum fittings with the DR7 composite spectrum reddened by 
the extinction curves from the SMC, the LMC2, the LMC and the MW  are also plotted.
The \ion{C}{4} BALs and the center of 2175 \AA~bump are marked with blue arrows.
(b). The best fit result of the parameterized extinction curve for J005335.59+085448.2 (Table \ref{tab1}).
(c). Histogram of fitted bump strength of the control sample for J005335.59+085448.2. 
The red line is the best fit Gaussian profile. The blue arrow indicates the strength of
the bump derived from the spectrum of J005335.59+085448.2. The three vertical dashed lines are the 3, 4, and
5$\sigma$ boundaries of the Gaussian. (d-g). The normalized flux of absorption lines in velocity space.}
\label{f1}
\end{figure*}

\figurenum{2}
\begin{figure*}[tbp]
\epsscale{1.} \plotone{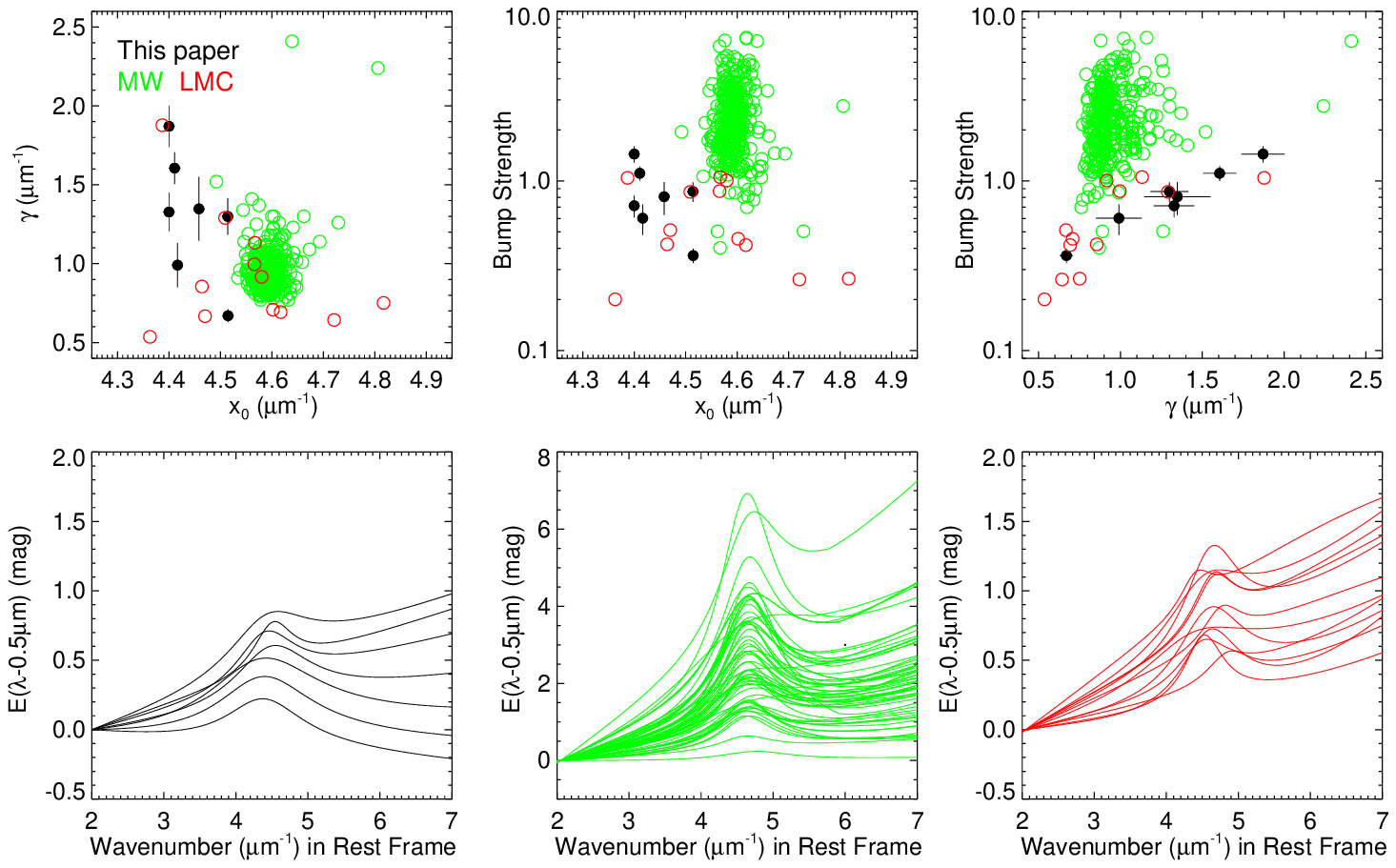}
\caption{Comparison of bump parameters (top panels) and relative extinction curves (bottom panels) with 2175 \AA~bumps for MW (Fitzpatrick \& Massa 2007) and LMC (Gordon et al. 2003) absorption curves. For visual clarity, the error bars on the MW and LMC bumps are not displayed.}\label{f2}
\end{figure*}

\subsection{Further confirmation of  bumps  in BAL quasars}
In order to confirm the existence of the EBBA bumps, we perform continuum fitting by reddening
the DR7 composite spectrum with different extinction curves, namely that from
the SMC, the LMC2 Supershell (hereafter LMC2), the LMC (Gordon et al. 2003) and the MW (Fitzpatrick \& Massa 2007).
The 2175 \AA~bump is absent in the SMC extinction curve and gradually increases from
the LMC2 extinction curve to the LMC and the MW extinction curves. There are two free parameters
in the continuum fitting: $E(B-V)$ and a normalization scale. The results of spectral fitting
are presented in Table \ref{tab1}.  We present the best fit results using different extinction curves by blue,
brown, pink and yellow curves for all the reddened quasar spectra in panel (a) of Figure \ref{f1}.
All objects show that the fitting reddened by a parameterized extinction curve is significantly 
better than those reddened by the SMC, LMC, LMC2 and MW extinction relations.

Considering that BAL quasars on average have  redder continua and 
stronger UV and optical \ion{Fe}{2} emission than non-BAL quasars
(e.g., Boroson \& Meyers 1992; Zhang et al. 2010),  a few of the strongest \ion{Fe}{2} and \ion{Fe}{3} emission objects
may be accidently identified as \2da~(Pitman et al. 2000). 
To check the validity of the detections of the EBBA~bumps, 
we  refit the spectra with the composite spectrum of HiBAL quasars from Reichard et al. (2003). 
The fitting results are shown by red lines (solid and dashed)
in panel (a) of Figure \ref{f1} and in Table \ref{tab2}. The HiBAL composite spectrum shows a redder continuum slope and
stronger UV \ion{Fe}{2} multiplets near the \ion{Mg}{2} regime than the DR7 quasar composite spectrum. 
The derived fitting parameters are somewhat different from the results using the DR7 composite spectrum.
The HiBAL fitting results show slightly stronger EBBA~bumps for all of the targets
except SDSS J135734.05+364005.1, confirming the detections of the bump. 
The maximal variability of the derived bump strengths reaches $\sim20\%$, 
indicating the level of uncertainties in the measured bump strengths due to the use of the different spectral template. 
Nevertheless, the anomalous absorption features are still clearly present.

To measure the significance level of the detections, which can also quantitatively
 assess the probability of false positives, 
we have compared the seven BAL quasar spectra with a control sample of 200 
quasar spectra  using the simulation technique developed by Jiang et al. (2010a,b).
This approed begins with the selection of a control sample of 200 quasar spectra with
 an $i$-band signal-to-noise ratio of $SNR\ge5$ 
in the range of $z_{em}-0.05<z<z_{em}+0.05$, where $z_{em}$ is the emission redshift of the studied quasar.
Then we fit each spectrum by reddening the DR7 composite quasar spectrum with a parameterized extinction
curve at the absorption redshift. The parameters $x_0$ and $\gamma$ in the parameterized extinction curve
are fixed to the best values fitting the EBBA bump of interest. 
The ``bump'' strengths, which are measured from all of the best fit bumps of the spectra in the control sample,
represent the random fluctuation of the quasar continuum and the variation of broad \ion{Fe}{2} emission multiplets
by assuming that no spectrum in the control sample possesses a real bump feature
 with the same width and peak position as the bump we detect in the candidates.
The distribution of ``bump'' strengths is expected to be a Gaussian profile as shown by the 
red curve in panel (c). In principle, the bumps with a significance level of $>3\sigma$\footnote{$\sigma$ is the standard deviation 
of the Gaussian profile.} are considered as real candidates.
All of the 2175 \AA~bumps reported in this paper 
are detected at $>4\sigma$ and five of them are $>5\sigma$, indicating real detections with a 
false positive probability of less than 0.1\%. 

\begin{deluxetable*}{lc cc cc cr cc cc}
\tabletypesize{\scriptsize}
\tablecaption{Best Fitted Parameters and Results of Fitting Spectra with Different Extinction Curves using the SDSS DR7 Quasar Composite Spectrum
\label{tab1} }
\tablewidth{0pt}
\tablehead{
\colhead{Name (SDSS J)}&
\colhead{$z$}&
\colhead{$c1$} &
\colhead{$c2$} &
\colhead{$c3$} &
\colhead{$x_0$} &
\colhead{$\gamma$} &
\colhead{$\chi^2_p$} &
\colhead{$\chi^2_{SMC}$}&
\colhead{$\chi^2_{LMC}$}&
\colhead{$\chi^2_{LMC2}$}&
\colhead{$\chi^2_{MW}$}
}
\startdata
005335.59+085448.2 &  1.78 & $1.81\pm0.01$ & $0.08\pm0.01$ & $0.72\pm0.02$& $4.54\pm0.01$ & $1.30\pm0.12$ & 1.06& 1.60&1.43&1.26&1.20\\
123852.97+420207.1 &  2.14 & $2.09\pm0.02$ & $0.02\pm0.01$ & $1.72\pm0.06$& $4.40\pm0.01$ & $1.87\pm0.13$ & 1.11& 1.39&1.53&1.57&1.49\\
125101.81+052441.5 &  1.92 & $2.33\pm0.02$ & $-0.02\pm0.01$& $1.13\pm0.04$& $4.41\pm0.01$ & $1.61\pm0.10$ & 1.06& 1.43&1.65&1.75&1.71\\
135734.05+364005.1 &  2.10 & $2.10\pm0.02$ & $0.13\pm0.01$ & $0.38\pm0.01$& $4.42\pm0.01$ & $0.99\pm0.14$ & 1.19& 1.40&1.24&1.22&1.30\\
155705.63+390805.3 &  2.26 & $1.65\pm0.02$ & $0.19\pm0.01$ & $0.69\pm0.03$& $4.46\pm0.01$ & $1.35\pm0.20$ & 1.53& 1.67&1.59&1.67&1.79\\
214653.22+004327.3 &  2.29 & $2.68\pm0.02$ & $-0.05\pm0.01$& $0.60\pm0.03$& $4.40\pm0.01$ & $1.33\pm0.12$ & 1.13& 1.33&1.62&1.97&2.00\\
232027.87+013428.2 &  1.84 & $1.14\pm0.01$ & $0.17\pm0.01$ & $0.15\pm0.01$& $4.51\pm0.01$ & $0.67\pm0.04$ & 1.11& 1.55&1.15&1.13&1.29\\
\hline
\\
HD 698 & & & & & 4.551$\pm$0.006 & 0.96$\pm$0.03 & & & & & \\            
HD 3191& & & & & 4.636$\pm$0.003 & 0.94$\pm$0.02 & & & & & \\            
Cr 463-5  & & & & & 4.610$\pm$0.008 & 1.08$\pm$0.05 & & & & &\\            
Cr 463-18 & & & & & 4.606$\pm$0.009 & 1.18$\pm$0.05 & & & & &\\            
BD +57 245& & & & & 4.561$\pm$0.005 & 0.89$\pm$0.03 & & & & & \\            
BD +57 252& & & & & 4.577$\pm$0.005 & 0.92$\pm$0.02 & & & & & \\            
BD +70 131& & & & & 4.577$\pm$0.011 & 1.11$\pm$0.05 & & & & &
\enddata
\tablenotetext{NOTE:}{$\chi^2_{?}$: Results of fitting the spectra by reddening the composite spectrum with the parameterized extinction curve and the average extinction curves of SMC, LMC, LMC2 and MW. 
The parameters for Galactic stars are adopted from Fitzpatrick \& Massa (2007). }
\end{deluxetable*}

\begin{deluxetable*}{r rr rr r}
\tabletypesize{\scriptsize}
\tablecaption{Best Fitted Parameters with Parameterized Extinction Curve using the HiBAL Quasar Composite Spectrum 
\label{tab2} }
\tablewidth{0pt}
\tablehead{
\colhead{Name (SDSS J)}&
\colhead{$c1$} &
\colhead{$c2$} &
\colhead{$c3$} &
\colhead{$x_0$} &
\colhead{$\gamma$}
}
\startdata
005335.59+085448.2 &  $2.09\pm0.02$ & $0.01\pm0.01$  & $1.00\pm0.03$& $4.51\pm0.01$ & $1.43\pm0.12$\\
123852.97+420207.1 &  $2.29\pm0.02$ & $-0.03\pm0.01$ & $2.32\pm0.07$& $4.40\pm0.01$ & $2.06\pm0.17$\\
125101.81+052441.5 &  $2.54\pm0.02$ & $-0.08\pm0.01$ & $1.51\pm0.04$& $4.41\pm0.01$ & $1.77\pm0.12$\\
135734.05+364005.1 &  $2.46\pm0.02$ & $0.06\pm0.01$  & $0.35\pm0.01$& $4.42\pm0.01$ & $1.09\pm0.14$\\
155705.63+390805.3 &  $1.92\pm0.02$ & $0.14\pm0.01$  & $0.77\pm0.03$& $4.46\pm0.01$ & $1.48\pm0.21$\\
214653.22+004327.3 &  $2.96\pm0.02$ & $-0.10\pm0.01$ & $0.72\pm0.03$& $4.41\pm0.01$ & $1.46\pm0.16$\\
232027.87+013428.2 &  $1.49\pm0.01$ & $0.10\pm0.01$  & $0.19\pm0.01$& $4.52\pm0.01$ & $0.73\pm0.14$
\enddata
\end{deluxetable*}

\section{Possible influence of broad iron emission}
To investigate quantitatively the possible influence of broad iron emission on the bump detection,
we constructed a simulated quasar sample with broad \ion{Fe}{2} emission multiplets at different intensity levels.
A total of 2,000 simulated quasar spectra  without 2175 \AA~bumps were 
created using the following parameters: their redshifts are fixed to $z = 2.0$; the simulated spectra cover
 the rest-frame wavelength range of $\sim1270-3070$ \AA, 
matching the wavelength coverage of bump features detected in the rest frame of the BAL quasars.
The simulated spectrum can be described by
$f(\lambda)=f_{\rm conti}(\lambda)+\beta f_{\rm Opt.FeII}(\lambda)+f_{\rm err}(\lambda)$,
where $f_{\rm conti}(\lambda)$ is the continuum plus broad emission lines, i.e., the SDSS DR7 composite 
quasar spectrum minus the broadened \ion{Fe}{2} template $f_{\rm Opt.FeII}(\lambda)$. 
$f_{\rm conti}(\lambda)$ and $f_{\rm Opt.FeII}(\lambda)$ are obtained using  the spectral component 
separation  technique introduced in Wang et al. (2009). The quantity $\beta$ is a free parameter, and the equivalent width
of the \ion{Fe}{2} emission, $EW_{\rm Opt.Fe II}$,
\footnote{The equivalent width is measured by integrating
 the broadened \ion{Fe}{2} template (Vestergaard \& Wilkes et al. 2001) over 2200 \AA~to 3050 \AA.} 
in the simulated sample ranges from 0.0 \AA~to 400.0 \AA.
$f_{\rm err}(\lambda)$ at each pixel is estimated from the flux density 
by the Gaussian error distribution, thus the signal-to-noise ratios of the simulated spectra 
can be matched the observed quasar sample.

This simulated quasar sample is used to search for the 2175 \AA~bump feature.
We set the bump width at $\gamma = 0.6 + 0.1*n$ ($n=0,1,...13$), $x_0=2175$ \AA,
while allow as the other parameters in Equation 1  vary freely. 
After the search, we obtained a false ``bump'' strength distribution on the two-dimensional
 grid  surface of the bump width and strength.  
The influence of \ion{Fe}{2} emission on bump detection gradually increases with
$EW_{\rm FeII}$ and decreases with the width of the 2175 \AA~bumps.
We derived the $3\sigma$ detection threshold for this simulated sample using the same procedure 
described earlier in this section and found that almost all ($>99\%$)
of the false ``bumps'' in the simulated spectra are under the 
3$\sigma$ detection threshold, except those quasars with strong iron emission
(e.g., 3 times above the average iron emission strength), which are extremely rare. 
 The most affected detections of the 2175 \AA\ absorbers are
those quasar-associated 2175 \AA~absorbers with narrow bumps.
For comparison, our seven candidates have bumps at  more than 4$\sigma$ significance levels. 
Moreover, the first six candidates have
bump widths of $\gamma\gtrsim1.00$, and the maximal value of their \ion{Fe}{2} emission equivalent 
width, $EW_{\rm FeII}$ of  $\sim 133$ \AA, is far less than the \ion{Fe}{2} emission ceiling value, 270 \AA, 
 of  the false detection required. Therefore, these detections are reliable. 
The last feature, in SDSS J232027.87+013428.2, has a narrow width of $\gamma = 0.67$ and an  $EW_{\rm FeII}$ of  261.9 \AA,
slightly less than the ceiling value of 270 \AA. Nevertheless, the detection level for SDSS J232027.87+013428.2
is at $\sim4\sigma$, and the bump detection in this system is also likely to be real.

\figurenum{3}
\begin{figure}[tbp]
\epsscale{1.3} \plotone{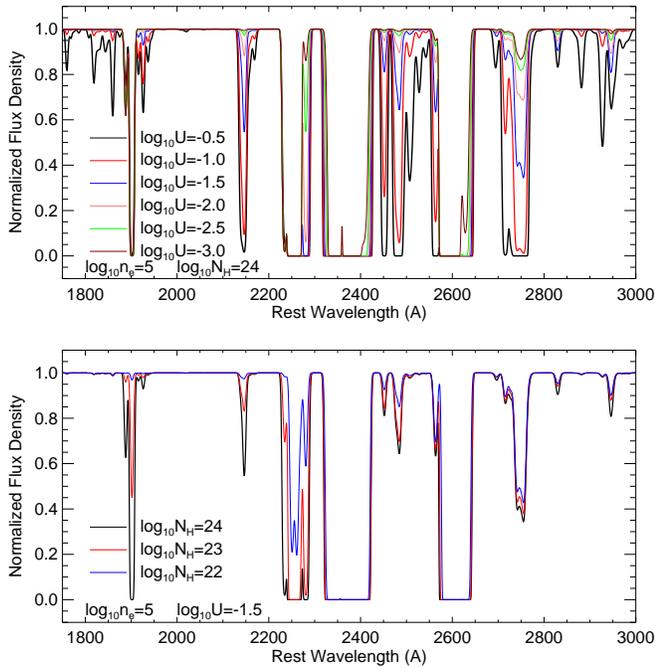}
\caption{Normalized spectra of \ion{Fe}{2} absorption lines in the 1750-3000 \AA~ region predicted by the model using CLOUDY.
In both panels, the blueshifted velocity of each \ion{Fe}{2} absorption line is set to 0 \kms, and their widths are 1000 \kms.}\label{f3}
\end{figure}

\begin{deluxetable}{lc cr cl }
\tabletypesize{\scriptsize}
\tablecaption{The properties of BALs 
\label{tab3} }
\tablewidth{0pt}
\tablehead{
\colhead{Name (SDSS J)}&
\colhead{Ty} &
\colhead{$V_{\rm min}$}&
\colhead{$V_{\rm max}$}&
\colhead{$D_{\rm max}$}&
\colhead{$BI$}
}
\startdata
005335.59+085448.2 &  Lo & 1885 & 18496 & $0.10\pm0.08$& $8701\pm19$\\ 
123852.97+420207.1 &  Hi & 2570 & 5025  & $0.11\pm0.07$& $1582\pm4 $\\ 
125101.81+052441.5 &  Lo & 1980 & 15589 & $0.00\pm0.08$& $7335\pm15$\\ 
135734.05+364005.1 &  Lo & 4743 & 11995 & $0.01\pm0.12$& $5104\pm12$\\ 
155705.63+390805.3 &  Lo & 2043 & 8020  & $0.00\pm0.09$& $3944\pm11$\\ 
214653.22+004327.3 &  Lo & 2444 & 11750 & $0.02\pm0.09$& $7137\pm8 $\\ 
232027.87+013428.2 &  Hi & 8455 & 18482 & $0.26\pm0.14$& $3649\pm16$ 
\enddata
\tablenotetext{NOTE:}{Ty: Subtypes of BALs: Hi for HiBAL, Lo for LoBAL.
$V_{\rm min}$ and $V_{\rm max}$: Minimum and maximum blueshifted velocity of \ion{C}{4} BAL troughs from the emission line (km s$^{-1}$).
$D_{\rm max}$: Residual flux at line bottom at the deepest part of \ion{C}{4} BAL troughs.
$BI$: Balnicity Index of \ion{C}{4} BAL troughs (km s$^{-1}$), defined by Gibson et al. (2009).}
\end{deluxetable}

\section{Are the EBBA bumps only an illusion of other unusual BALs?}

\ion{C}{4} BALs are detected in  the spectra of all of the seven objects, while
 \ion{Si}{4} BALs appear in all spectra except SDSS J123852.97+420207.1.
We normalized the observed spectra using the best-fit models (green solid lines in Figure \ref{f1}),
and measured \ion{C}{4} absorption lines in the normalized spectra (panel (d)).
The “balnicity index” (BI) of \ion{C}{4} BALs are listed in Table \ref{tab3}.
Table \ref{tab3} also shows the minimal and maximal velocities of \ion{C}{4} BAL troughs from the emission line,
the typical widths are several thousand \kms, while the maximal velocity reaches approximatively twenty thousand \kms.
{\rev ``Balnicity" indexes are several thousands (much larger than 0) 
and residual fluxes in the deepest part of the troughs are close to zero, 
indicate the strong absorption of \ion{C}{4} BALs.}

We classify the seven BALs in our sample as either HiBALs and LoBALs based on the  modified ``absorption index (AI)'' introduced 
 in Zhang et al. (2010). The AI is extracted from an observed quasar spectrum after removing 
absorption components from those low ionized species such as \ion{Mg}{2} and
\ion{Al}{3}.  The unabsorbed model spectrum is  obtained through  multicomponent fitting of the \ion{Mg}{2} and
\ion{Al}{3} spectral regimes. Specifically,  we adopted  a single power law for the continuum, 
two or three Gaussians for the broad \ion{Mg}{2} and \ion{Al}{3} lines, and the broadened UV \ion{Fe}{2} emission multiplets
(also see Lu et al. 2008) to fit the broad \ion{Mg}{2} and \ion{Al}{3} lines to derive their AI. 
The \ion{Mg}{2} or \ion{Al}{3} BAL troughs
should fall at least 10\% below the model spectrum in a contiguous velocity interval of at least 1600 \kms 
to be considered to be real.
Based on the AI measurements, five objects are classified as LoBALs of \ion{Mg}{2} and \ion{Al}{3}
(e.g., \ion{Mg}{2} BALs shown in panel (g) of Figure \ref{f1}). The five of BALs are marked 
with `Lo' in Table \ref{tab3}, while the two remaining objects
are defined as HiBALs due to lack of   \ion{Mg}{2} and \ion{Al}{3} absorption troughs.

Since all seven objects in our sample are BAL quasars, a question is naturally raised: 
Are the EBBA bumps only an illusion  
caused by other unusual BALs, such as \ion{Fe}{2} and/or \ion{Fe}{3}?
This confusion could occur numerous iron absorption systems cover the same broad wavelength regime as
the bump, and some lines even coincidently lie near the bump peak position.
To investigate the existence of \ion{Fe}{2} absorption,
we calculated a series of models using the photoionization code CLOUDY (c10.00; Ferland et al. 1998) 
with different ionization parameters and hydrogen column densities, 
adopting the ionizing continuum from Mathews \& Ferland (1987) 
and an electron density of $n_e = 10^5$ cm$^{-3}$. In addition, the default 16-level Fe atom is used.
In the top panel of Figure \ref{f3}, the ionization parameter ($log_{10}U$) is varied between
 -0.5 to -3.0 with a dex step of 0.5,
and the highest column density is chosen to be $N_{\rm H} = 10^{24}$ cm$^{-2}$.
In the bottom panel of Figure \ref{f3}, the highest column density ($log_{10}N_{\rm H}$) is set
between  22 and 24 with a dex step of 1.0,
and the ionization parameter is chosen to be $U = 10^{-1.5}$.
The calculations suggest that the iron absorption troughs around 2400 \AA~ (\ion{Fe}{2} UV 2) 
and 2600 \AA~ (\ion{Fe}{2} UV 1) are the strongest and the most pervasive. 
In panel (f) of Figure \ref{f1}, the normalized flux of \ion{Fe}{2} UV1 
is compared with \ion{C}{4} and/or \ion{Mg}{2} BALs.
The influence of \ion{Fe}{2} BALs can be ruled out from visual examination
due to the lack of detections of the commonly associated  \ion{Fe}{2} BALs in these seven spectra.

For \ion{Fe}{3} BALs, Hall et al. (2002) reported one  \ion{Fe}{3}-dominant
 BAL quasar (SDSS J221511.94-004549.9)
and three possible candidates (SDSS J014905.28-011404.9, J081024.75+480615.5 and J21441.43-000137.9).
\ion{Fe}{3} absorption is produced by \ion{Fe}{3} UV48 $\lambda\lambda\lambda$2062.21,2068.90,2079.65
and \ion{Fe}{3} UV34 $\lambda\lambda\lambda$1895.46,1914.06,1926.30.
These four objects show unprecedented line ratios, e.g., 
 with stronger absorption in \ion{Al}{3} than \ion{Mg}{2},
and FeLoBALs with weak or no \ion{Fe}{2} but with strong \ion{Fe}{3}, which is 
 even stronger than \ion{Mg}{2}.
However, the troughs of \ion{Fe}{3} UV34, \ion{Fe}{3} UV48, \ion{Al}{3} and
\ion{Mg}{2} have the same blueshifted velocity  and similar velocity structure.
Although the wavelength regime is coincidentally covered by the 2175 \AA~bump,
the troughs of \ion{Fe}{3} UV34, \ion{Fe}{3} UV48 are located well shortward of the peak position of
the 2175 \AA~bump. At the same time, all objects reported in this paper do not show the
strong broad absorption features except the known \ion{Al}{3} BALs around 2000 \AA~(e.g., \ion{Fe}{3} UV48 shown 
in panel (e) of Figure \ref{f1}). Therefore, we conclude that the detected 2175 \AA~bumps 
 are unlikely to be caused by the illusion of \ion{Fe}{3} BALs.

\section{Summary and Discussion}

In this paper,  for the first time, we report the discovery of excess broad band absorption near 2250 \AA~
in the spectra of seven BAL quasars. 
These objects were selected from  the  SDSS-III BOSS DR10 quasar
database using the same \ion{Mg}{2} absorption identification method reported in Jiang et al (2011). 
These EBBA features are detected at a high confidence level (4-5$\sigma$).
They have broader widths at smaller wavenumbers, 
and their strengths are much weaker than those of the MW bumps on average, but similar to the LMC. 
The EBBA feature may be related to the 2175 \AA~bump seen in the Local Group (MW, LMC, SMC) 
and may be a counterpart of the 2175 \AA~bump under different conditions in the early Universe.
If true, they are the only candidates showing both  BALs and 2175 \AA-related bumps in the same spectra.
Moreover, HiBALs are present in all of the seven quasar spectra while loBALs appear in  five objects.

 BALs are the imprints of outflows driven by super-massive black holes
residing in the centers of AGNs  (Weymann et al. 1991).
Previous studies indicate that LoBALs may be early progenitors of normal AGNs and quasars
(e.g., Farrah et al. 2007; Urrutia et al. 2009), at the very end of an extreme starburst phase. 
The $Spitzer$ spectra of some FeLoBALs, the extreme subclass of LoBALs, have further shown significant
signatures of dust and polycyclic aromatic hydrocarbon emissions (Farrah et al. 2010).
Previous statistical works indicate that
 about 15\% of quasars show BALs, and another $\sim$15\% of BAL quasars have detected LoBALs
(Weymann et al. 1991; Hewett \& Foltz 2003; Reichard et al.2003; Trump et al. 2006; Gibson et al. 2009; Zhang et al. 2010).
When we searched for \2da~in the SDSS-III/BOSS DR10 quasar spectra, we identified a total of
 25 quasar-associated 2175 \AA\ absorber candidates. Therefore, 
about 28\% of the quasar-associated \2da~show  BALs, of which 71\%  (or five of seven) show  LoBALs.
Both fractions are substantially higher than those in the overall quasar sample,
and the true fractions should be  even  higher because BAL quasars are dimmed in the observed optical wavelengths 
by dust reddening in the \2da~and can be easily missed in a magnitude-limited sample.
This situation suggests that there is a strong correlation between occurences of LoBALs and 2175 \AA~bumps.
The discovery of BAL quasars with 2175 \AA-related bumps~may offer a new possibility
to study the connection among the central engine, the outflow of a quasar, and the dust in the host galaxy.

A probable global picture about the early evolution of quasars and their host galaxies appears from this 
study. The persistent processes of nucleosynthesis of the element carbon and formation  of dust
occur throughout the evolution of stars and galaxies, preparing the material that eventually produces 
the broad absorption near 2175 \AA.
Dust grains with proper size and composition do not exist in all quasars.
For the quasar-associated 2175 \AA~ or 2175 \AA-related bumps, the dusty materials are associated with the AGN environments,
probably the ISM of the host galaxy, and/or the dusty envelope layer near the quasar nucleus.
In the early stage of evolution, the center of AGNs/quasars are encapsulated.
When the central black holes are ignited, they release vast quantities of radiant energy,
including high-energy ionizing photons (Rees et al. 1984).
The radiation photoionizes and heats the surrounding gas, and drives strong outflows of gas and dust.
(Davidson \& Netzer 1979; Weymann et al. 1991).
When the line of sight to the continuum source intersects the outflow, BALs are produced in the spectrum.
Outflows can efficiently shield the high-energy photons from the central engine and extend the survival period of dust.
Meanwhile, the energy feedback from outflows clears the surrounding space of the core
(Silk \& Rees 1998; Fabian 1999; Di Matteo et al. 2005).
Accompanied by the large energy feedback,
the dusty envelope layer is destroyed and cleared gradually.
In this process, low-ionization ions are produced and survive in two regions, the dusty envelope layer
and the core or the trailing end of outflows, because of the shielding by outflows.
This appears to naturally explain the high fraction of LoBALs in the 2175 \AA~extinction quasars identified in this 
paper.

\acknowledgments
Many thanks to  Dr. Xueguang Zhang and Dr. Tinggui Wang for very useful discussions
and Dr. Xiheng Shi for the help with CLOUDY computations.
This work is supported by the University of Florida and also Chinese Natural Science Foundation by NSFC-11203021,
National Basic Research Program of China (973 Program, 2013CB834905) 
and the SOC program (CHINARE2012-02-03).  P.J. acknowledges supports from NSFC with grants NSFC-11233002 and NSFC-11203022.
This work has made use of the data obtained by SDSS-III.
Funding for SDSS-III has been provided by the Alfred P. Sloan Foundation, the Participating Institutions, the National Science Foundation, and the U.S. Department of Energy Office of Science. The SDSS-III web site is http://www.sdss3.org/.

SDSS-III is managed by the Astrophysical Research Consortium for the Participating Institutions of the SDSS-III Collaboration including the University of Ari- zona, the Brazilian Participation Group, Brookhaven National Laboratory, Carnegie Mellon University, Uni- versity of Florida, the French Participation Group, the German Participation Group, Harvard University, the Instituto de Astrofisica de Canarias, the Michigan State/Notre Dame/JINA Participation Group, Johns Hopkins University, Lawrence Berkeley National Labora- tory, Max Planck Institute for Astrophysics, Max Planck Institute for Extraterrestrial Physics, New Mexico State University, New York University, Ohio State University, Pennsylvania State University, University of Portsmouth, Princeton University, the Spanish Participation Group, University of Tokyo, University of Utah, Vanderbilt Uni- versity, University of Virginia, University of Washington, and Yale University.%
%


\clearpage

\figurenum{1}
\begin{figure*}[bp]
\epsscale{1.} \plotone{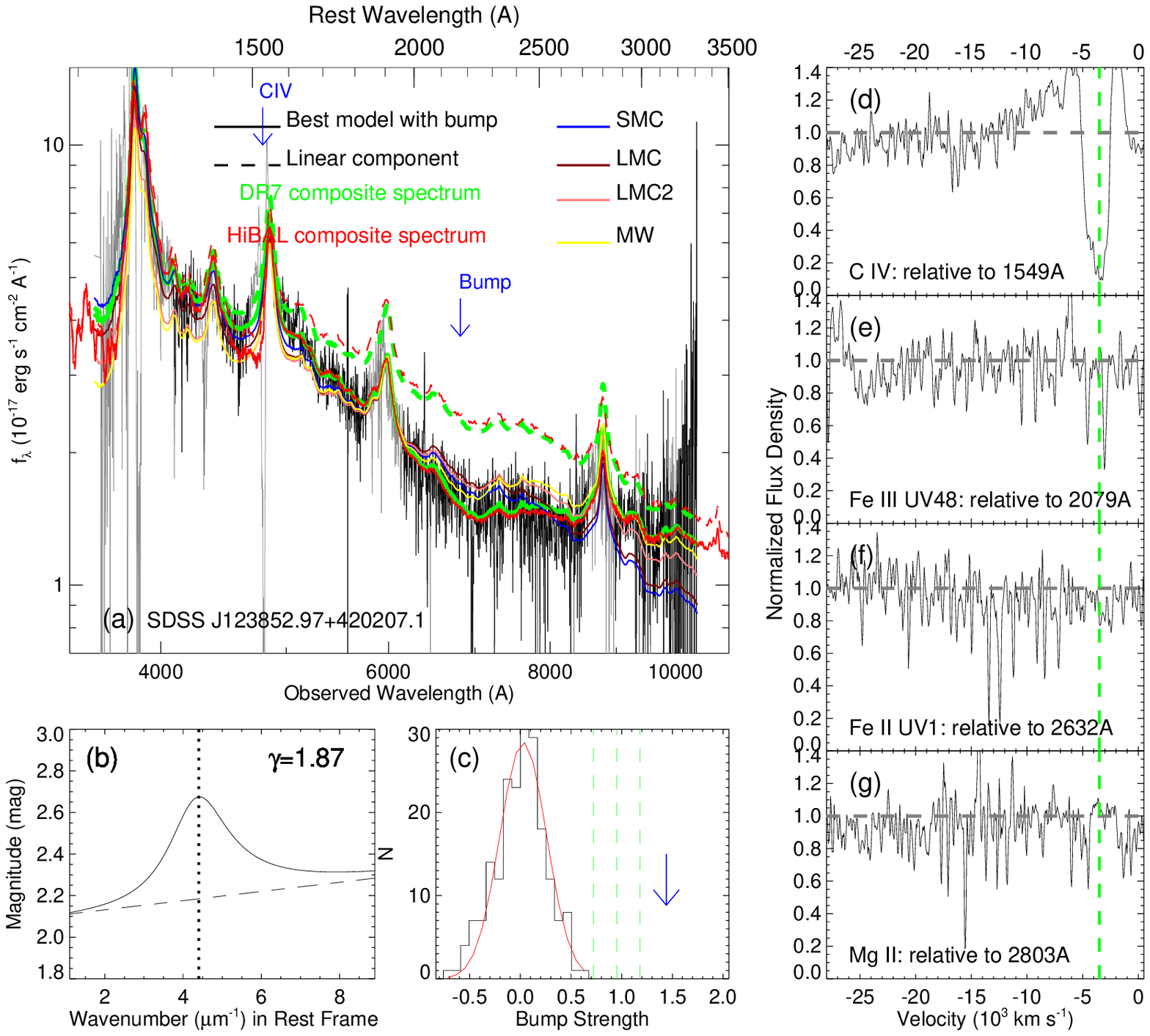}
\caption{(Continued) The best fitted extinction model and absorption lines for J123852.97+420207.1.}\label{f1}
\end{figure*}
\clearpage

\figurenum{1}
\begin{figure*}[bp]
\epsscale{1.} \plotone{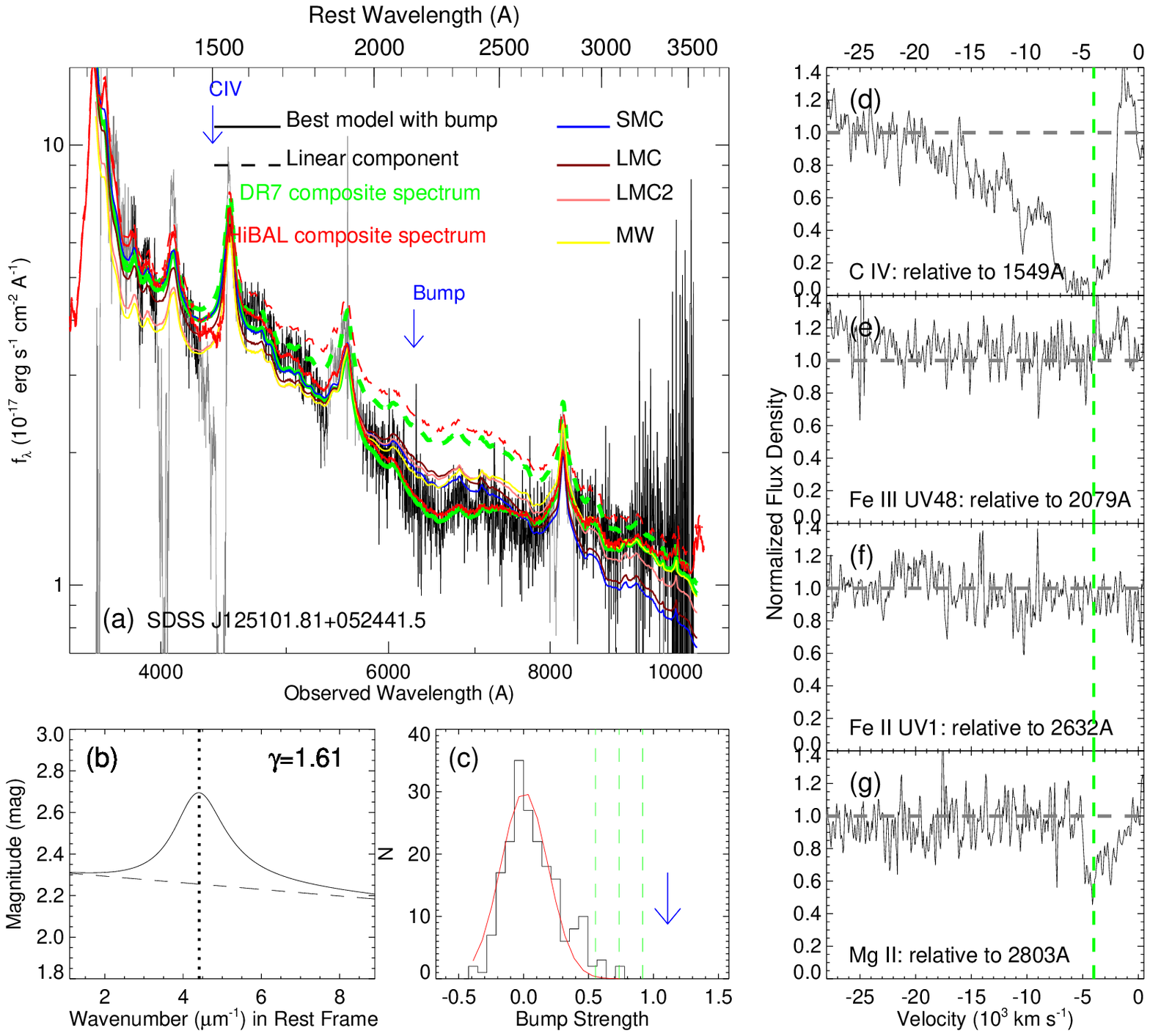}
\caption{(Continued) The best fitted extinction model and absorption lines for J125101.81+052441.5.}\label{f1}
\end{figure*}
\clearpage

\figurenum{1}
\begin{figure*}[bp]
\epsscale{1.} \plotone{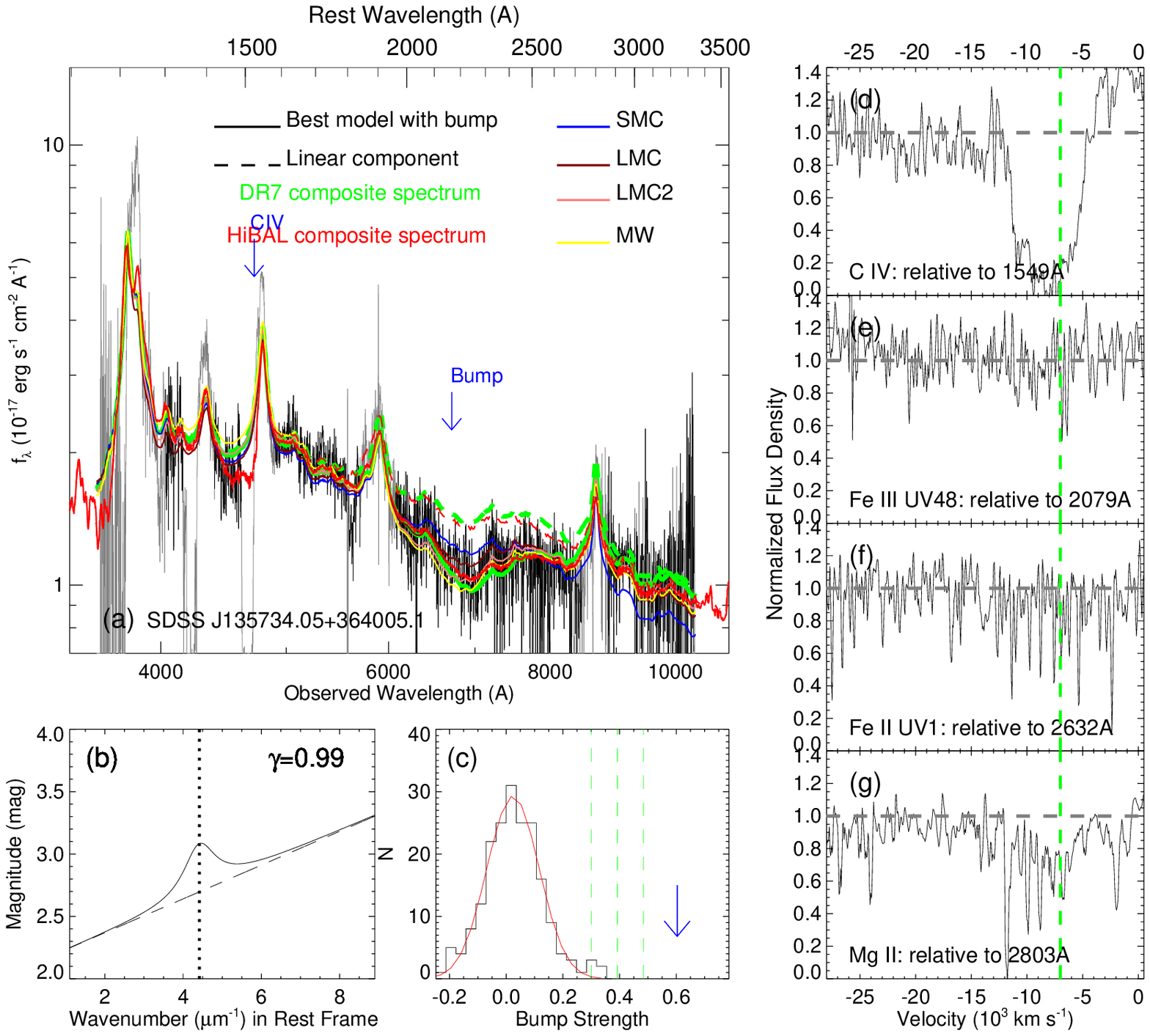}
\caption{(Continued) The best fitted extinction model and absorption lines for J135734.05+364005.1.}\label{f1}
\end{figure*}
\clearpage

\figurenum{1}
\begin{figure*}[bp]
\epsscale{1.} \plotone{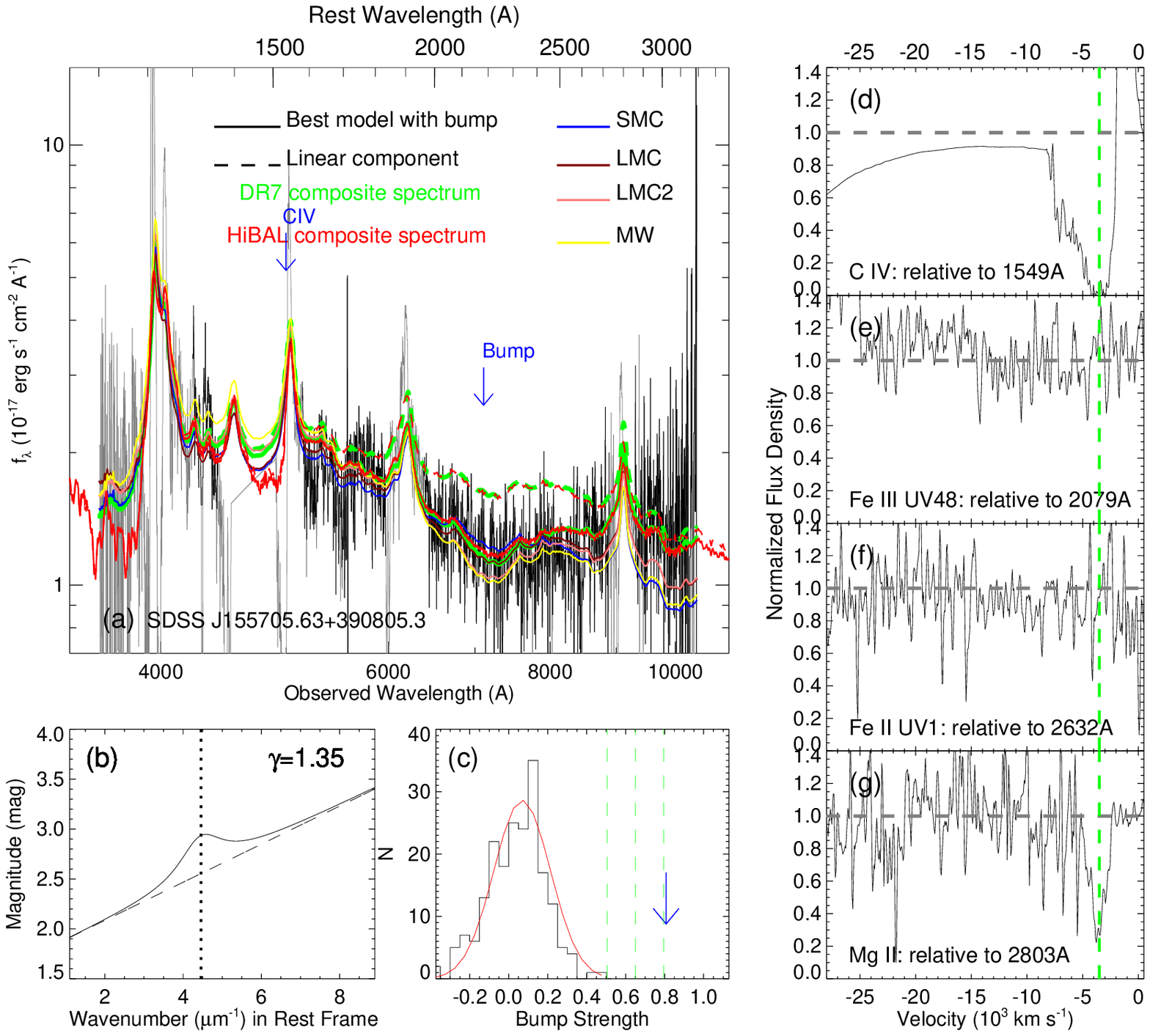}
\caption{(Continued) The best fitted extinction model and absorption lines for J155705.63+390805.3.}\label{f1}
\end{figure*}
\figurenum{1}
\clearpage

\figurenum{1}
\begin{figure*}[bp]
\epsscale{1.} \plotone{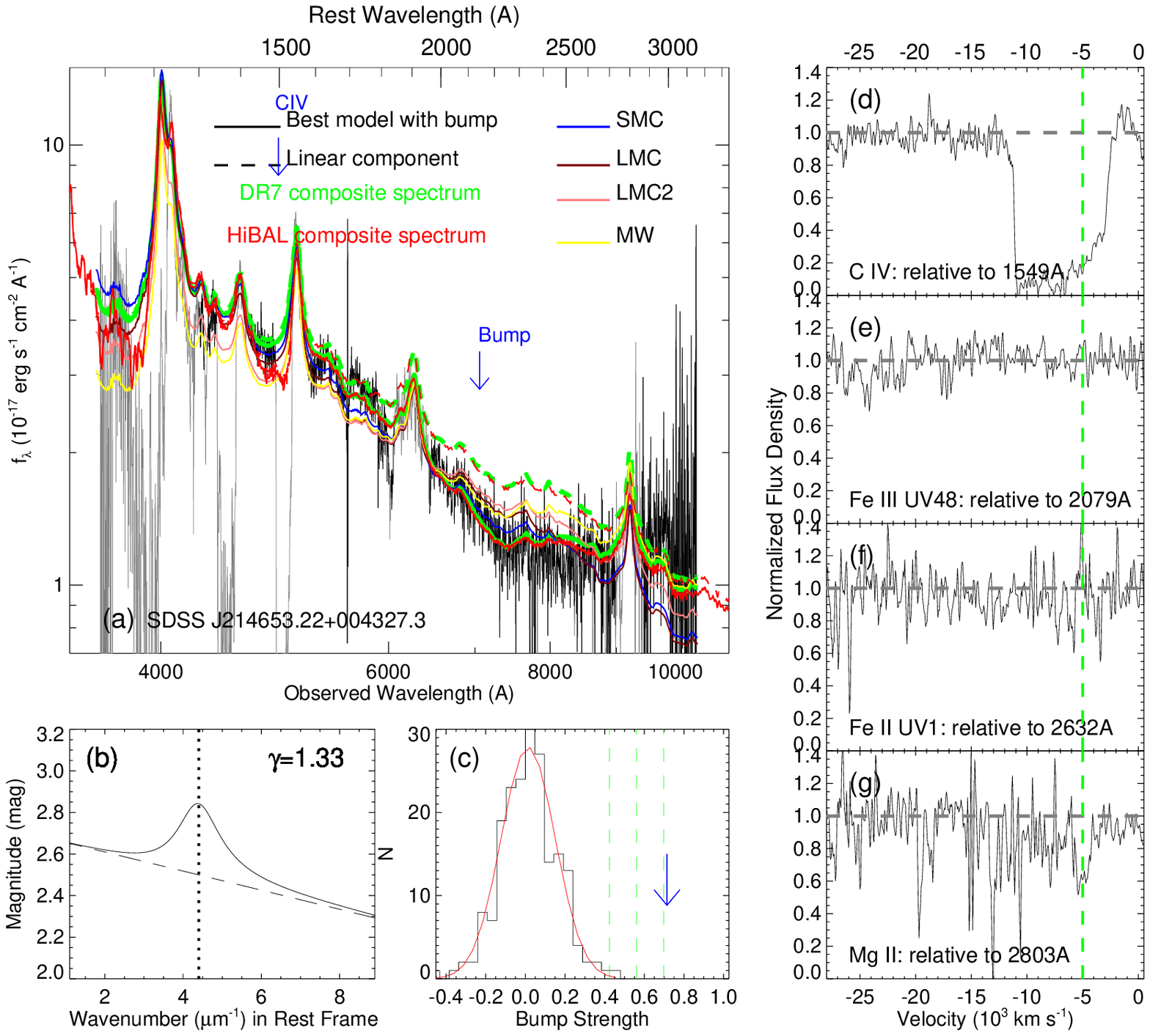}
\caption{(Continued) The best fitted extinction model and absorption lines for J214653.22+004327.3.}\label{f1}
\end{figure*}
\clearpage

\figurenum{1}
\begin{figure*}[bp]
\epsscale{1.} \plotone{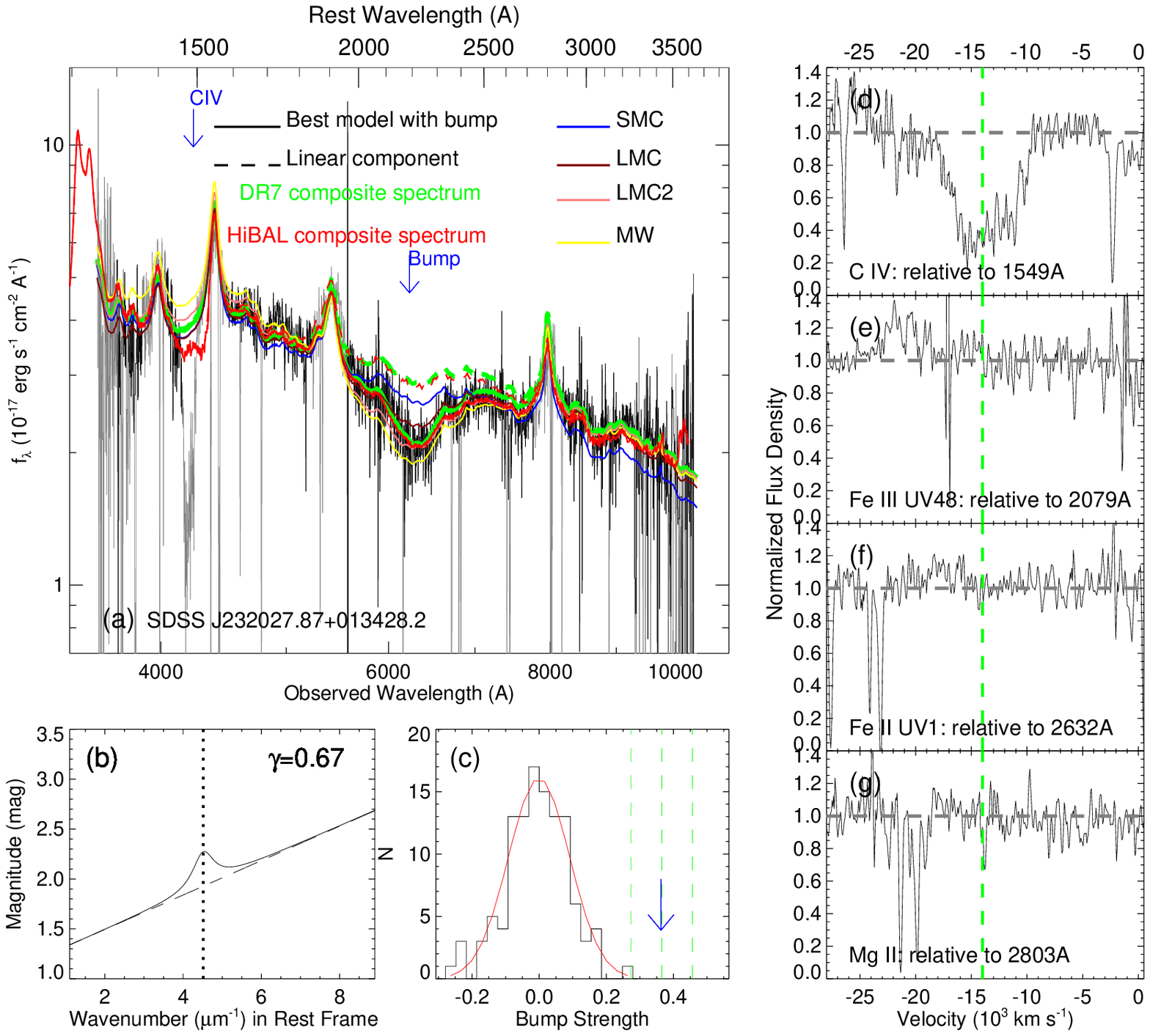}
\caption{(Continued) The best fitted extinction model and absorption lines for J232027.87+013428.2.}\label{f1}
\end{figure*}
\clearpage


\begin{thebibliography}{}

\bibitem[Ahn et al.(2014)]{2014ApJS..211...17A} Ahn, C.~P., Alexandroff, R., Allende Prieto, C., et al.\ 2014, \apjs, 211, 17
\bibitem[Antonuccio-Delogu \& Silk(2010)]{2010ASPC..427..343A} Antonuccio-Delogu, V., \& Silk, J.\ 2010, Accretion and Ejection in AGN: a Global View, 427, 343
\bibitem[Brandt et al.(2000)]{2000ApJ...528..637B} Brandt, W.~N., Laor, A., \& Wills, B.~J.\ 2000, \apj, 528, 637
\bibitem[Boroson \& Meyers(1992)]{1992ApJ...397..442B} Boroson, T.~A., \& Meyers, K.~A.\ 1992, \apj, 397, 442
\bibitem[Davidson \& Netzer(1979)]{1979RvMP...51..715D} Davidson, K., \& Netzer, H.\ 1979, Reviews of Modern Physics, 51, 715
\bibitem[Dawson et al.(2013)]{2013AJ....145...10D} Dawson, K.~S., et al.\ 2013, \aj, 145, 10
\bibitem[Di Matteo et al.(2005)]{2005Natur.433..604D} Di Matteo, T., et al. \ 2005, \nat, 433, 604
\bibitem[Draine(2003)]{2003ARA&A..41..241D} Draine, B.~T.\ 2003, \araa, 41, 241
\bibitem[Eisenstein et al.(2011)]{2011AJ....142...72E} Eisenstein, D.~J., et al.\ 2011, \aj, 142, 72
\bibitem[Fabian(1999)]{1999MNRAS.308L..39F} Fabian, A.~C.\ 1999, \mnras, 308, L39
\bibitem[Farrah et al.(2007)]{2007ApJ...662L..59F} Farrah, D., et al.\ 2007, \apjl, 662, L59
\bibitem[Farrah et al.(2010)]{2010ApJ...717..868F} Farrah, D., et al.\ 2010, \apj, 717, 868
\bibitem[Ferland et al.(1998)]{1998PASP..110..761F} Ferland, G.~J., et al.\ 1998, \pasp, 110, 761 
\bibitem[Fitzpatrick(1989)]{1989IAUS..135...37F} Fitzpatrick, E.\ 1989, Interstellar Dust, 135, 37
\bibitem[Fitzpatrick \& Massa(1990)]{1990ApJS...72..163F} Fitzpatrick, E.~L., \& Massa, D.\ 1990, \apjs, 72, 163
\bibitem[Fitzpatrick \& Massa(2007)]{2007ApJ...663..320F} Fitzpatrick, E.~L., \& Massa, D.\ 2007, \apj, 663, 320
\bibitem[Fynbo et al.(2013)]{2013ApJS..204....6F} Fynbo, J.~P.~U., Krogager, J.-K., Venemans, B., et al.\ 2013, \apjs, 204, 6 
\bibitem[Gibson et al.(2009)]{2009ApJ...692..758G} Gibson, R.~R., Jiang, L., Brandt, W.~N., et al.\ 2009, \apj, 692, 758
\bibitem[Green \& Mathur(1996)]{1996ApJ...462..637G} Green, P.~J., \& Mathur, S.\ 1996, \apj, 462, 637
\bibitem[Gordon et al.(2003)]{2003ApJ...594..279G} Gordon, K.~D., Clayton, G.~C., et al.\ 2003, \apj, 594, 279
\bibitem[Gunn et al. (2006)]{2006AJ...131...2332} Gunn, J. E., et al.\ 2006, \aj, 131, 2332
\bibitem[Hall et al.(2002)]{2002ApJS..141..267H} Hall, P.~B., Anderson, S.~F., et al.\ 2002, \apjs, 141, 267
\bibitem[Hamann \& Sabra(2004)]{2004ASPC..311..203H} Hamann, F., \& Sabra, B.\ 2004, AGN Physics with the Sloan Digital Sky Survey, 311, 203
\bibitem[Hewett \& Foltz(2003)]{2003AJ....125.1784H} Hewett, P.~C., \& Foltz, C.~B.\ 2003, \aj, 125, 1784
\bibitem[Jiang et al.(2010)]{2010ApJ...720..328J} Jiang, P., Ge, J., Prochaska, J.~X., et al.\ 2010a, \apj, 720, 328
\bibitem[Jiang et al.(2010)]{2010ApJ...724.1325J} Jiang, P., Ge, J., Prochaska, J.~X., et al.\ 2010b, \apj, 724, 1325
\bibitem[Jiang et al.(2011)]{2011ApJ...732..110J} Jiang, P., Ge, J., Zhou, H., et al.\ 2011, \apj, 732, 110
\bibitem[Jiang et al.(2013)]{2013AJ....145..157J} Jiang, P., Zhou, H., Ji, T., et al.\ 2013, \aj, 145, 157 
\bibitem[Lu et al.(2008)]{2008ApJ...680..858L} Lu, H., Wang, T., Yuan, W., et al.\ 2008, \apj, 680, 858
\bibitem[Mathews \& Ferland(1987)]{1987ApJ...323..456M} Mathews, W.~G., \& Ferland, G.~J.\ 1987, \apj, 323, 456 
\bibitem[Murray \& Chiang(1995)]{1995ApJ...454L.105M} Murray, N., \& Chiang, J.\ 1995, \apjl, 454, L105 
\bibitem[Noterdaeme et al.(2009)]{2009A&A...503..765N} Noterdaeme, P., et al.\ 2009, \aap, 503, 765
\bibitem[Noterdaeme et al.(2014)]{2014A&A...566A..24N} Noterdaeme, P., Petitjean, P., P{\^a}ris, I., et al.\ 2014, \aap, 566, A24
\bibitem[P{\^a}ris et al.(2014)]{2014A&A...563A..54P} P{\^a}ris, I., Petitjean, P., Aubourg, {\'E}., et al.\ 2014, \aap, 563, AA54 
\bibitem[Pitman et al.(2000)]{2000PASP..112..537P} Pitman, K.~M., et al.\ 2000, \pasp, 112, 537
\bibitem[Rees(1984)]{1984ARA&A..22..471R} Rees, M.~J.\ 1984, \araa, 22, 471
\bibitem[Reichard et al.(2003)]{2003AJ....125.1711R} Reichard, T.~A., Richards, G.~T., et al.\ 2003, \aj, 125, 1711
\bibitem[Ross et al.(2012)]{2012ApJS..199....3R} Ross, N.~P., et al.\ 2012, \apjs, 199, 3
\bibitem[Savage \& Mathis(1979)]{1979ARA&A..17...73S} Savage, B.~D., \& Mathis, J.~S.\ 1979, \araa, 17, 73
\bibitem[Silk \& Rees(1998)]{1998A&A...331L...1S} Silk, J., \& Rees, M.~J.\ 1998, \aap, 331, L1
\bibitem[Smee et al.(2013)]{2013AJ....146...32S} Smee, S.~A., Gunn, J.~E., 
Uomoto, A., et al.\ 2013, \aj, 146, 32
\bibitem[Sprayberry \& Foltz(1992)]{1992ApJ...390...39S} Sprayberry, D., \& Foltz, C.~B.\ 1992, \apj, 390, 39
\bibitem[Srianand et al.(2008)]{2008MNRAS.391L..69S} Srianand, R., et al.\ 2008, \mnras, 391, L69
\bibitem[Stecher(1965)]{1965ApJ...142.1683S} Stecher, T.~P.\ 1965, \apj, 142, 1683
\bibitem[Tolea et al.(2002)]{2002ApJ...578L..31T} Tolea, A., Krolik, J.~H., \& Tsvetanov, Z.\ 2002, \apjl, 578, L31
\bibitem[Trump et al.(2006)]{2006ApJS..165....1T} Trump, J.~R., et al.\ 2006, \apjs, 165, 1
\bibitem[Urrutia et al.(2009)]{2009ApJ...698.1095U} Urrutia, T., Becker, R.~H., et al.\ 2009, \apj, 698, 1095
\bibitem[Vestergaard \& Wilkes(2001)]{2001ApJS..134....1V} Vestergaard, M., \& Wilkes, B.~J.\ 2001, \apjs, 134, 1
\bibitem[Voit(1992)]{1992MNRAS.258..841V} Voit, G.~M.\ 1992, \mnras, 258, 841
\bibitem[Wang et al.(2004)]{2004ApJ...609..589W} Wang, J., et al.\ 2004, \apj, 609, 589
\bibitem[Wang et al.(2005)]{2005pgqa.conf..331W} Wang, J., Ge, J., Hall, P.~B., Prochaska, J.~X., 
\& Li, A.\ 2005, IAU Colloq.~199: Probing Galaxies through Quasar Absorption Lines, 331 
\bibitem[Wang et al.(2009)]{2009ApJ...707.1334W} Wang, J.-G., Dong, X.-B., 
Wang, T.-G., et al.\ 2009, \apj, 707, 1334 
\bibitem[Wang et al.(2012)]{2012ApJ...760...42W} Wang, J.-G., et al.\ 2012, \apj, 760, 42
\bibitem[Weymann et al.(1991)]{1991ApJ...373...23W} Weymann, R.~J., et al. \ 1991, \apj, 373, 23
\bibitem[Wild et al.(2006)]{2006MNRAS.367..211W} Wild, V., Hewett, P.~C., \& Pettini, M.\ 2006, \mnras, 367, 211
\bibitem[York et al.(2000)]{2000AJ....120.1579Y} York, D.~G., et al.\ 2000, \aj, 120, 1579
\bibitem[York et al.(2006)]{2006MNRAS.367..945Y} York, D.~G., et al.\ 2006, \mnras, 367, 945
\bibitem[Zhang et al.(2010)]{2010ApJ...714..367Z} Zhang, S., Wang, T.-G., Wang, H., et al.\ 2010, \apj, 714, 367
\bibitem[Zhang et al.(2014)]{2014ApJ...786...42Z} Zhang, S., Wang, H., Wang, T., et al.\ 2014, \apj, 786, 42 
\bibitem[Zhou et al.(2010)]{2010ApJ...708..742Z} Zhou, H., Ge, J., Lu, H., et al.\ 2010, \apj, 708, 742
%
\end{thebibliography}
\end{document}